\documentclass[prb,twocolumn,showpacs]{revtex4}
\usepackage[dvips]{graphicx}
\usepackage[tight,normal]{subfigure}
\usepackage{amsmath}
\usepackage{units}
\usepackage{amssymb}
\usepackage{amsfonts}
\usepackage{ae}

\begin{document}
%
\title{Superconducting pumping of nanomechanical vibrations}
\author{Gustav Sonne}
\affiliation{Department of Physics, University of Gothenburg, SE-412 96 G\"oteborg, Sweden}
\author{Leonid Y. Gorelik}
\email{gorelik@chalmers.se}
\affiliation{Department of Applied Physics, Chalmers University of Technology, SE-412 96 G\"oteborg, Sweden}
\author{Sergei I. Kulinich}
\affiliation{Department of Applied Physics, Chalmers University of Technology, SE-412 96 G\"oteborg, Sweden}
\affiliation{B.I. Verkin Institute for Low Temperature Physics and Engineering, 47 Lenin Avenue, 611 03 Kharkov, Ukraine}
\author{Robert I. Shekhter}
\affiliation{Department of Physics, University of Gothenburg, SE-412 96 G\"oteborg, Sweden}
\author{Mats Jonson}
\affiliation{Department of Physics, University of Gothenburg, SE-412 96 G\"oteborg, Sweden}
\affiliation{School of Engineering and Physical Sciences, Heriot-Watt University, Edinburgh EH14 4AS, Scotland, UK}
\date{\today}
%
\begin{abstract}
We demonstrate that a supercurrent can pump energy from a battery
that provides a voltage bias into nanomechanical vibrations. Using a
device containing a nanowire Josephson weak link as an example we
show that a nonlinear coupling between the supercurrent and a static
external magnetic field leads to a Lorentz force that excites
bending vibrations of the wire at resonance conditions. We also
demonstrate the possibility to achieve more than one regime of
stationary nonlinear vibrations and how to detect them via the
associated dc Josephson currents and we discuss possible
applications of such a multistable nanoelectromechanical dynamics.
\end{abstract}
\pacs{74.45.+c, 74.50.+r, 74.78.Na, 75.47.De}
\maketitle

Coupling of electronic and mechanical degrees of freedom on the
nanometer length scale is the basic phenomenon behind the
functionality of nanoelectromechanical (NEM) systems. Such a coupling
can be mediated either by electrical charges or currents.
Single-electron tunneling (SET) devices with movable islands or gate
electrodes employ Coulomb forces to achieve capacitive
\cite{Naik,Knobel} and shuttle NEM coupling \cite{Gorelik1998,
Shekhter2007}, where the latter involves both capacitive forces and
charge transfer. Devices containing current carrying parts, on the
other hand, will achieve NEM coupling through magnetic-field induced
Lorentz forces. Focusing on the latter mechanism, a simple estimate
shows that for a gold nanowire suspended over a few micrometer long
trench, the mechanical displacement due to typical currents of order
\unit[100]{nA} in magnetic fields of order \unit[0.01]{T} can be as
large as one nanometer.  Such displacements can crucially affect the
performance of mesoscopic devices.

In this Letter we will explore a possible scenario for how highly
{\em nonlinear} nanoelectromechanical effects can arise if the
magnetic-field induced electromotive force caused by the mechanical
motion of a conducting wire strongly perturbs the flow of current
through it. Devices which contain superconductors, with their known
extreme sensitivity to external electric fields, are the best
candidates to achieve such strong effects and superconducting
quantum interference devices (SQUID's) that incorporate a
nanomechanical resonator are particularly interesting. Significant
research has recently been performed in this direction (see e.g
Ref.~\onlinecite{Buks, Buks2,Blencowe,Zhou}), by using a coupling
between the SQUID dynamics and the resonator's mechanical vibrations
due to the constraint set by the flux quantization phenomenon.
Here we will consider the new possibility for NEM coupling that
occurs if the nanomechanical element is an integral part of the
superconducting weak link. In this case the NEM vibrations directly
affect the Cooper pair tunneling and significantly modify the
properties of the link. With a voltage biased weak link it becomes
possible to pump nanomechanical vibrations in the Cooper pair
tunneling region. As we will show below the result is a peculiar
nonlinear NEM dynamics that affect both the supercurrent flow and
the nanomechanical vibrations in novel ways.
\begin{figure}[h]
\begin{center}
\includegraphics[width=0.4\textwidth]{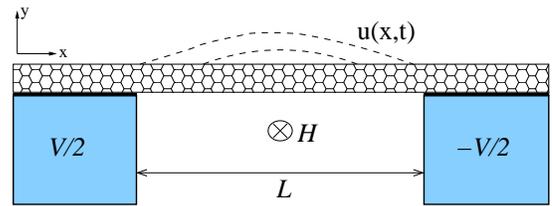}
\caption{(Color online) Sketch of system considered. A nanowire is
  suspended between two superconducting electrodes separated by a
  trench of length $L$. When the system is biased by a voltage $V$,
  the Lorentz force caused by the coupling of the Josephson current
  and a transverse magnetic field, $H$, induces wire vibrations
  described by the coordinate $u(x,t)$. The nonlinear coupling leads
  to a multistability of the system resulting in different dc
  Josephson current regimes (see text).}
\label{picture}
\end{center}
\end{figure}

The Hamiltonian describing the electronic subsystem in the specific
model system shown in Fig.~\ref{picture} reads,
\begin{gather}
\hat{\mathcal{H}}=\int\textrm{d}x\hat{\Psi}^{\dagger}(x)\left(\hat{\mathcal{H}}_0+\hat{\mathcal{H}}_{\Delta}\right)\hat{\Psi}(x)\notag\\
\hat{\mathcal{H}}_0=-\frac{\hbar^2}{2m}\sigma_z\left(\frac{\partial }{\partial x}-\sigma_z\frac{ieHu(x,t)}{\hbar}\right)^2+\sigma_zU(x)\\
\hat{\mathcal{H}}_{\Delta}=\Delta(x)\left(\sigma_x\cos\phi(t)+\text{sign}(x)\sigma_y\sin\phi(t)\right)\notag\,,
\label{Hamil}
\end{gather}
where $\Psi^{\dagger}(x)$ [$\Psi(x)$] are two-component
Nambu-spinors and $\sigma_i$ are the Pauli matrices in Nambu space
\footnote{The Zeeman term in the Hamiltonian is small for the magnetic
fields of interest here and is ignored.}. The deflection of the tube
is given by $u(x,t)=u(x)a(t)$, where $u(x)$ is the normalized
(dimensionless) profile of the fundamental bending mode and $a(t)$
determines its amplitude; other modes are less important and will be
ignored. The potential $U(x)$ describes the barrier between the
nanowire and the bulk superconducting electrodes, where the gap
parameter is $\Delta(x)=\Delta_0\theta(2\vert x\vert -L)$ with
$\Delta_0\sim$\unit[10]{meV}.  The phase difference across the
junction due to the bias voltage $V$ is $\phi(t)=2eVt/\hbar$.

A convenient gauge transformation, see Ref. \onlinecite{Shekhter} for a similar
analysis, shifts the vector potential induced by the nanotube
deflection from the kinetic part of the Hamiltonian to the phase
difference between the leads, so that
$\phi(t)\rightarrow\varphi(t)=\phi(t)-a(t)4eH\int_0^{L/2}u(x)\text{d}x/\hbar$.
In the adiabatic limit, $\hbar D\dot{\varphi}(t)\ll \Delta_0$, with
$D$ the transparency of the barriers, one can then evaluate the
fixed-phase ground state energy of the electronic subsystem \cite{Bagwell} as
$E(\varphi)=-\Delta_0[1-D\sin^2(\varphi/2)]^{1/2}$, and
find that the force exerted on the wire, $F = -\partial
E(\varphi(a))/\partial a$, is proportional to the Josephson current
$j=(2e/\hbar)\partial E(\varphi)/\partial \varphi$.  The resulting
effective equation of motion for the nanowire vibrating in its
fundamental bending mode describes a forced nonlinear oscillator with
damping. In terms of the dimensionless coordinate
$Y(t)=[4eLH/\hbar]a(t)$ one finds in the low transparency limit, $D\ll
1$, the result
\begin{subequations}\label{eomboth}
\begin{gather}
\ddot{Y}+\tilde{\gamma}\dot{Y}+Y=\epsilon\sin(\varphi)\label{eom}\\
\dot{\varphi}=\widetilde{V}-\dot{Y}\label{phaseevo}\,.
\end{gather}
\end{subequations}
Here, $\tilde{\gamma}=\gamma/m\omega$ is a dimensionless damping
coefficient, while $\epsilon=8eL^2H^2j_c/(m\hbar\omega^2)$ is the
amplitude and $\dot{\varphi}$ the frequency of the driving force with
$\widetilde{V}=2eV/(\hbar\omega)$, $\omega$ the mechanical eigenfrequency,
$m$ the mass of the nanowire, $j_c=D\Delta_0e/(2\hbar)$ the critical
current and time $t$ measured in units of $1/\omega$. In \eqref{eom},
the driving force on the nanowire is naturally interpreted as the
Lorentz force due to the coupling between the Josephson current and
the magnetic field, which, due to the confined geometry of the charge
carriers in the nanowire, is responsible for depositing energy from
the electronic to the mechanical subsystem. According to
\eqref{phaseevo}, the phase difference $\varphi$ between the leads
evolves in time under the influence of both the bias voltage and the
electromotive force induced by the motion of the wire in the static
magnetic field.

Multiplying \eqref{eom} with $\dot{Y}$ and averaging over time we
find (using the definition of the Josephson current above) that in
the stationary regime the dc current through the system is
\begin{equation}
j_{dc}=\frac{\gamma \langle
\dot{a}(t)^2\rangle}{V}=\frac{\gamma\hbar^2\omega^2\langle\dot{Y}(t)^2\rangle}{16e^2L^2H^2
V} \label{dc} ,
\end{equation}
where $\langle...\rangle$ denotes time-averaged quantities.

To proceed with our analysis we consider the specific case of a
single-wall carbon nanotube wire of diameter \unit[1]{nm} suspended
over a length $L\sim$ \unit[1]{$\mu$m}. With $j_c\sim$ \unit[100]{nA}
\cite{Kasumov} one then finds that $\epsilon\sim$
\unit[3$\times$10$^{-3}$] in a magnetic field of $H\sim$
\unit[20]{mT}. Since $\tilde{\gamma}=1/Q$, where the quality factor
$Q\sim 1000$ \cite{Witkamp,Sazonova}, both $\epsilon$ and
$\tilde{\gamma}$ may be considered small, $\epsilon,\tilde{\gamma}\ll
1$.

\begin{figure}[h]
\begin{center}
\hspace*{-.01in}
\subfigure[
$\,\,\,\epsilon=0.001$]{\label{fig:Q100x0}\includegraphics[width=0.24\textwidth]{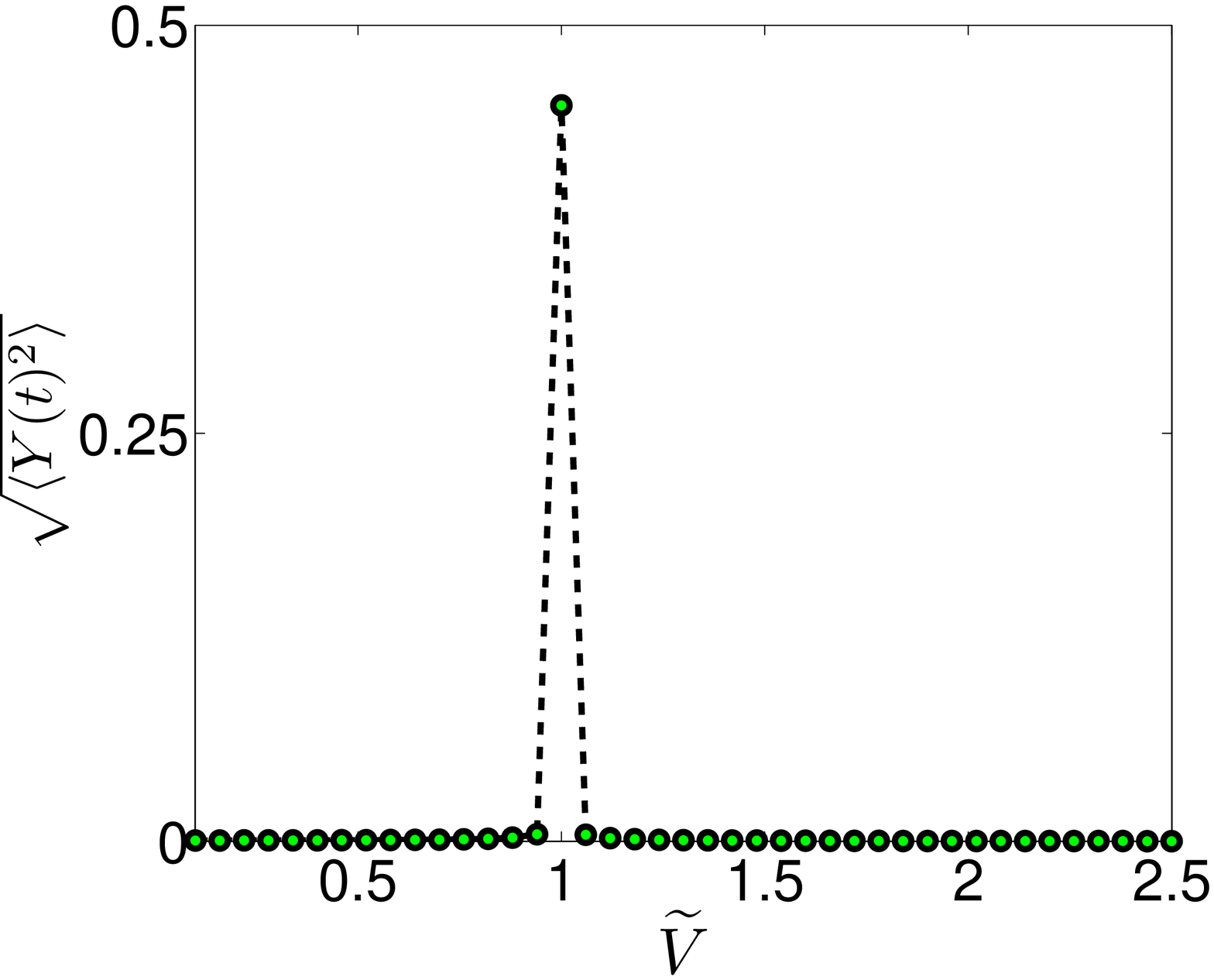}}\hspace*{-.03in}
\subfigure[
$\,\,\,\epsilon=0.01$]{\label{fig:Q1000x0}\includegraphics[width=0.24\textwidth]{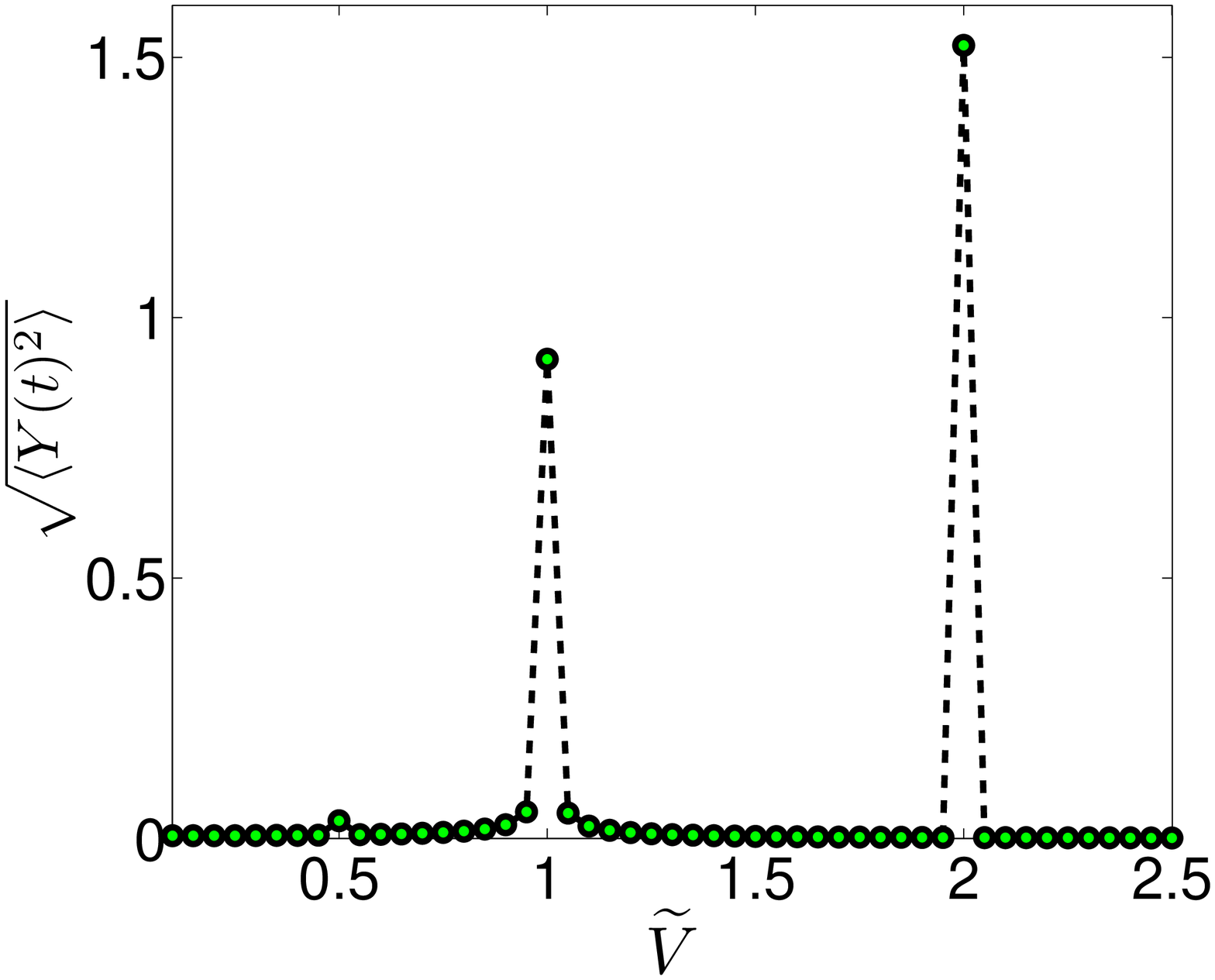}}
\caption{(Color online) Time average of the rms nanowire deflection
  coordinate $Y(t)$ from a numerical simulation of Eq.~\eqref{eomboth}
  as a function of bias voltage $\tilde V$ for different force
  parameters $\epsilon$ ($\tilde\gamma=0.001$).}
\label{firstimage}
\end{center}
\end{figure}

Numerical simulations of the nanowire dynamics using the equation of
motion \eqref{eomboth} with initial conditions $Y(0)=\dot{Y}(0)=0$
show distinct resonance peaks in the vibration amplitude at integer
values of $\widetilde{V}$.  Figure~\ref{firstimage}, e.g. shows peaks at
$\tilde{V}$=1 and 2 (as well as a small peak at $\widetilde{V}=1/2$),
where the onset of the $\widetilde{V}$=2 peak depends on the ratio
$\epsilon/\tilde\gamma$
\footnote{As will be discussed elsewhere, peaks may appear also for
$\tilde{V}=3,4...$ depending on the initial conditions used.}.
For small vibration amplitudes, when one can expand
$\sin(\widetilde{V}t-Y)$ to linear order in $Y$, these results can readily
be attributed to a direct resonance at $\tilde{V}=1$ and a parametric
resonance at $\widetilde{V}=2$. In this limit there is a resemblance
between the resonances in our system and the familiar Fiske effect in
Josephson junctions coupled to an electromagnetic resonator
\cite{Barone}. However, as can be seen in Fig.~\ref{fig:Q1000x0}, the
oscillation amplitude is too large for the linear approximation to
hold if the driving force is large, $\epsilon > \tilde{\gamma}$. As
will be shown below, the resonances in this nonlinear regime are
significantly different from those of the Fiske effect and demonstrate
a variety of unusual peculiarities which could be useful for device
applications.

To analyze the nonlinear regime in the vicinity of the resonance peaks
it is convenient to apply perturbation theory and expand in the small
parameters $\epsilon$ and $\tilde{\gamma}$.  With this in mind, it is
useful to write the dimensionless deflection coordinate $Y(t)$ of the
nanowire as
\begin{equation}
Y(t)=\sqrt{I_n(t)}\sin\left(\frac{\widetilde{V}t}{n}+\frac{\chi_n(t)}{n}\right),
\label{ansatz}
\end{equation}
where the amplitude $A_n(t)\equiv\sqrt{I_n(t)}$ and phase 
$\chi_n(t)$ vary slowly in time; $\dot{I}_n(t),\dot{\chi}_n(t)\approx
\tilde{\gamma},\epsilon\ll 1$. Substituting this Ansatz into
\eqref{eomboth} and integrating over the fast oscillations one gets
two coupled equations for $I_n(t)$ and $\chi_n(t)$,
\begin{subequations}\label{solution}
\begin{gather}
\dot{I}_n=-\tilde{\gamma} I_n-2\epsilon n J_n(\sqrt{I_n})\sin\chi_n\label{solution1}\\
\dot{\chi}_n=-\delta-2\epsilon n J'_n(\sqrt{I_n})\cos\chi_n\,.\label{solution2}
\end{gather}
\end{subequations}
Here, $J_n$ are Bessel functions of order $n$,
$J'_n(\sqrt{I_n})=dJ_n(\sqrt{I_n})/dI_n$ and
$\delta=\widetilde{V}-n$. Stationary nonlinear oscillation regimes can now
be found by studying the stationary points of \eqref{solution} in
terms of the system parameters.

We start our analysis by considering the case of exact resonance,
$\delta=0$, for which Eq.~(\ref{solution}) guarantees that a
stationary solution given by $I_n=0$ always exists.  However, since
$J_n(x)\sim x^{n}$ for small $x$, one immediately finds that this
solution is unstable for $n=1$ (resonance excitation), while for $n=2$
(parametric excitation) it is only unstable if
$\epsilon>2\tilde{\gamma}$.  This turns out to be the main difference
between the two resonance types; the corresponding finite-amplitude
stationary regimes are qualitatively very similar. In the following
analysis we will therefore focus on the parametric resonance at $n=2$,
and omit the index $n$ on amplitudes, phases and Bessel functions.
\begin{figure}[h]
\begin{center}
\hspace*{-.17in}\subfigure[]{\label{Besselfig}\includegraphics[width=0.26\textwidth]{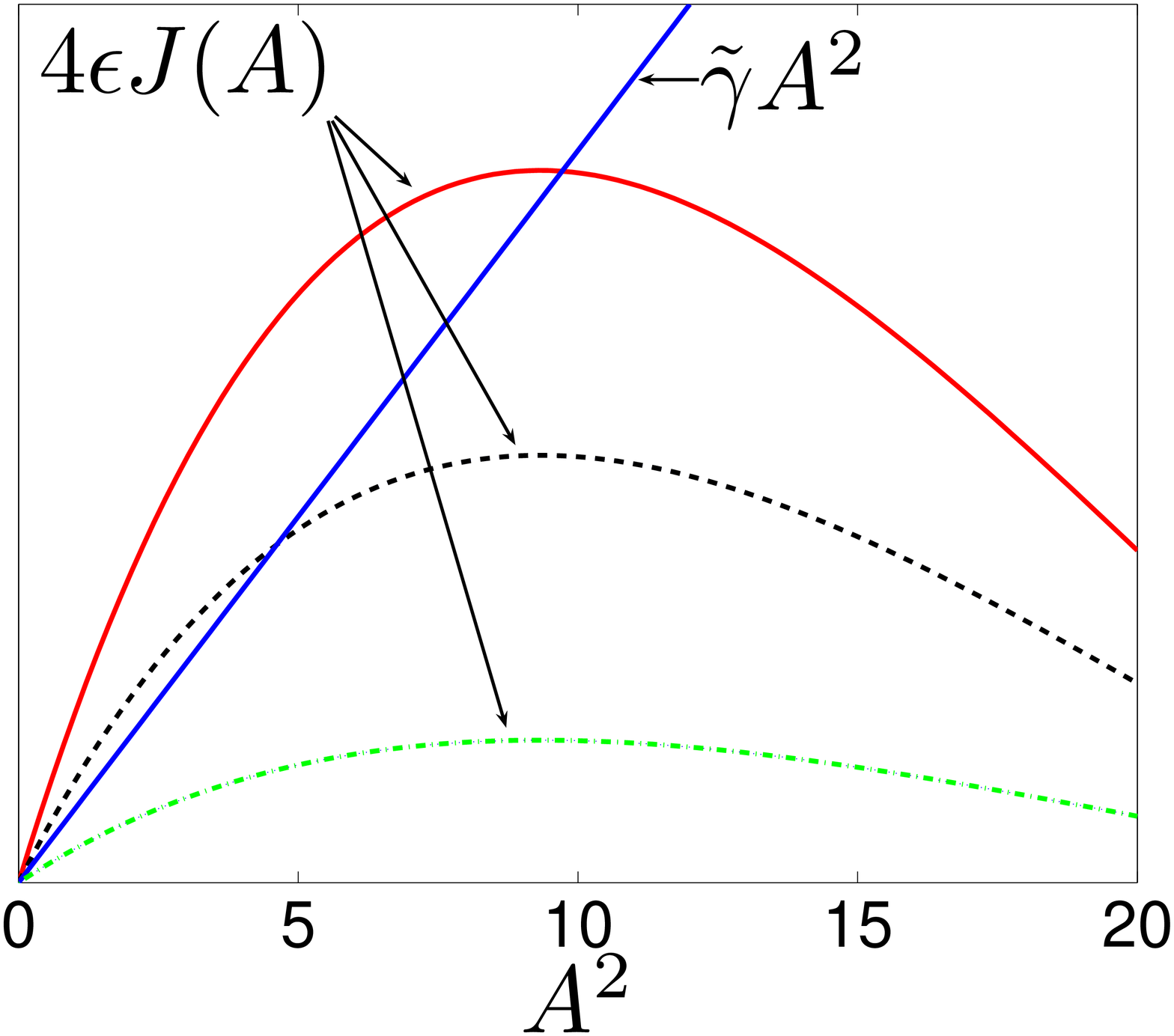}}\hspace*{-.14in}
\subfigure[]{\label{eplot}\includegraphics[width=0.26\textwidth]{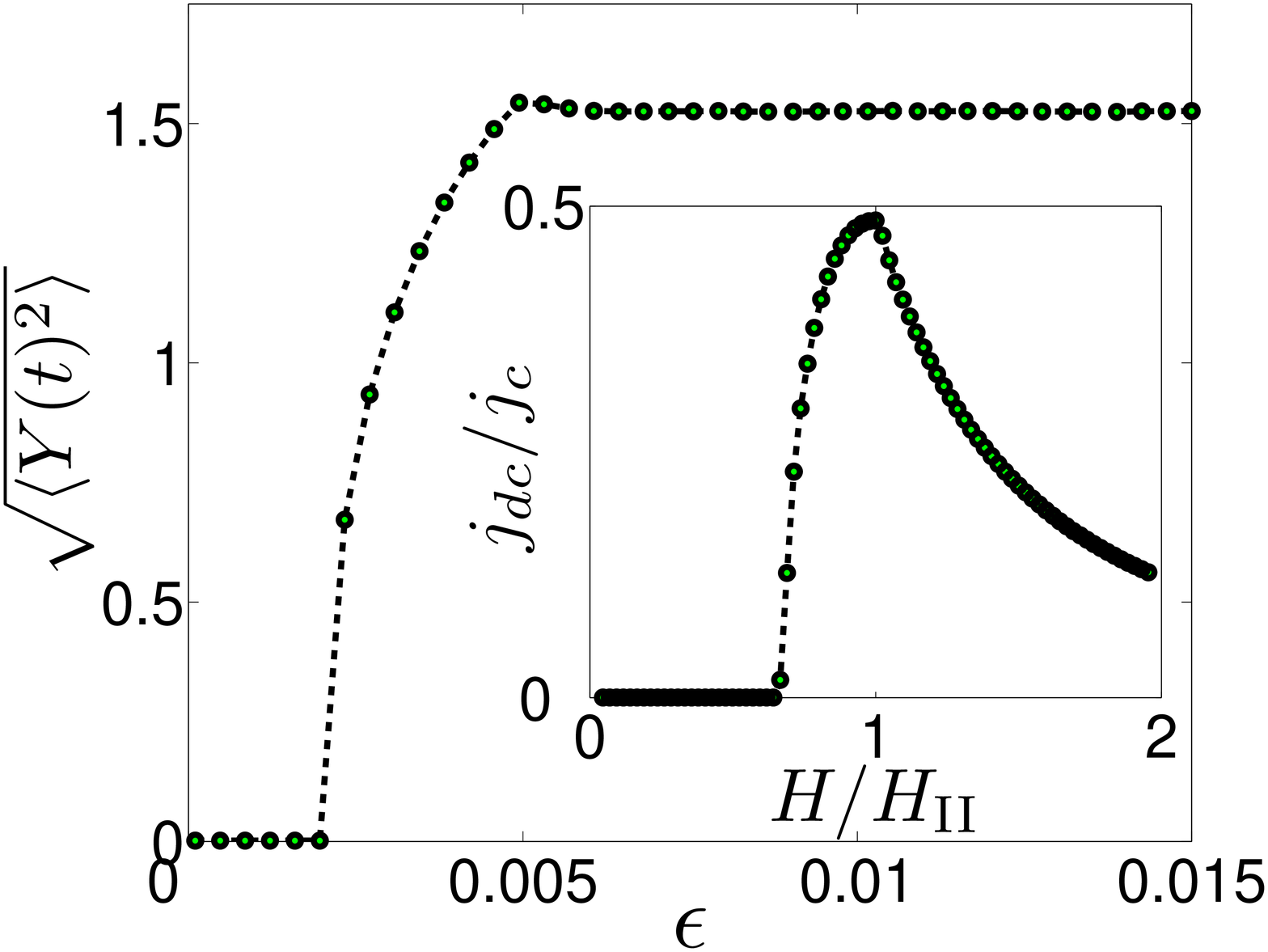}}
\caption{(Color online) (a) Plots for solution of
  Eq.~\eqref{solution1} with $\dot{I}_n=0$ for $\chi_n=3\pi/2$
  (type-$\mathrm{I}$ regime) and $\epsilon=0.005$ (red solid/dark gray),
  $\epsilon=0.003$ (black dashed) and $\epsilon=0.001$ (green dashed
  dotted/light gray). Crossings with the straight line correspond to stationary
  points. (b) Numerical solution on resonance of Eq.~\eqref{eomboth}
  for the time-averaged rms nanowire deflection coordinate as a
  function of $\epsilon$. Inset shows corresponding plot for dc
  current as a function of magnetic field. $\delta=0$ and
  $\tilde{\gamma}=0.001$ throughout.}
\label{epsilonvary}
\end{center}
\end{figure}

From Eq.~\eqref{solution2} it is evident that exactly on resonance a
finite-amplitude stationary regime may be realized either by fixing
the phase, $\cos\chi=0$, or the amplitude, $J'(A)=0$. We will refer to
these different regimes as type $\mathrm{I}$ and type $\mathrm{II}$.
From \eqref{solution1} it follows that a type-$\mathrm{I}$ stationary
point exists for any $\epsilon>\epsilon_{\mathrm{I}}\equiv
2\tilde{\gamma}$. The oscillation amplitude is implicitly given by the
equation
$\tilde{\gamma}/\epsilon=4J(A_{\mathrm{I}})/A_{\mathrm{I}}^{2}$, which
always has a solution in the relevant range of parameters, see
Fig.~\ref{Besselfig} \footnote{Equation (\ref{solution}a) with
$\dot{I}_n=0$ may have more than one solution if the ratio
$\epsilon/{\tilde\gamma}$ is large. If so, we will always refer to the
one with smallest amplitude.}.  Furthermore, type-$\mathrm{II}$
stationary points corresponding to fixed-amplitude oscillations,
$A_{\mathrm{II}}=A_0$, where $J'(A_{0})=0$, only exists if
$\epsilon>\epsilon_{\mathrm{II}}\equiv\tilde{\gamma}A^2_{0}/4J(A_{0})$.
In this case there exists two stationary points of equal amplitude,
$A_{\mathrm{II}}^{\pm}=A_0$, but different phases,
$\chi_{\mathrm{II}}^{\pm}=3\pi/2\pm\arccos(\epsilon_{\mathrm{II}}/\epsilon)$
\footnote{Which type-$\mathrm{II}$ stationary point that corresponds
  to the true solution depends on the initial conditions.}.

A stability analysis shows that the type-I stationary point is stable
if $\epsilon<\epsilon_{\mathrm{II}}$, but unstable (a saddle point)
for $\epsilon>\epsilon_{\mathrm{II}}$. The type-II stationary points,
on the other hand, are always stable if they exist, i.e.  when
$\epsilon>\epsilon_{\mathrm{II}}$.  This means that if one increases
$\epsilon$, by turning up the magnetic field, the nanotube vibration
amplitude will be zero (to an accuracy of order
$\epsilon,\tilde{\gamma}$) until
$\epsilon\sim\epsilon_{\mathrm{I}}$. As $\epsilon$ is varied from
$\epsilon_{\mathrm{I}}$ to $\epsilon_{\mathrm{II}}$ the amplitude
increases from 0 to $A_0$, where it saturates as we increase the
magnetic field further. This analysis, which also explains the onset
of the second peak in Fig.~\ref{fig:Q1000x0}, has been fully confirmed
by numerically solving the equation of motion \eqref{eomboth} for the
vibration amplitude at varying values of $\epsilon$, as shown in
Fig.~\ref{eplot}. The inset shows the dc current as a function of
magnetic field, where $H_{\mathrm{II}}$ is defined from
$\epsilon_{\mathrm{II}}\propto H_{\mathrm{II}}^2$ as above. As the dc
current scales as $j_{dc}\propto\langle\dot{Y}(t)^2\rangle/H^{2}$ one
finds that the current initially grows with increasing magnetic field
strength, pumping energy into the nanoscale vibrations, but falls off
as $1/H^{2}$ once $H>H_{\mathrm{II}}$ and the vibration amplitude has
saturated at $A_0$.

\begin{figure}[h]
\begin{center}
\hspace*{-.08in}
\subfigure[$\,\,\,\delta=0$]{\label{exact}\includegraphics[width=0.25\textwidth]{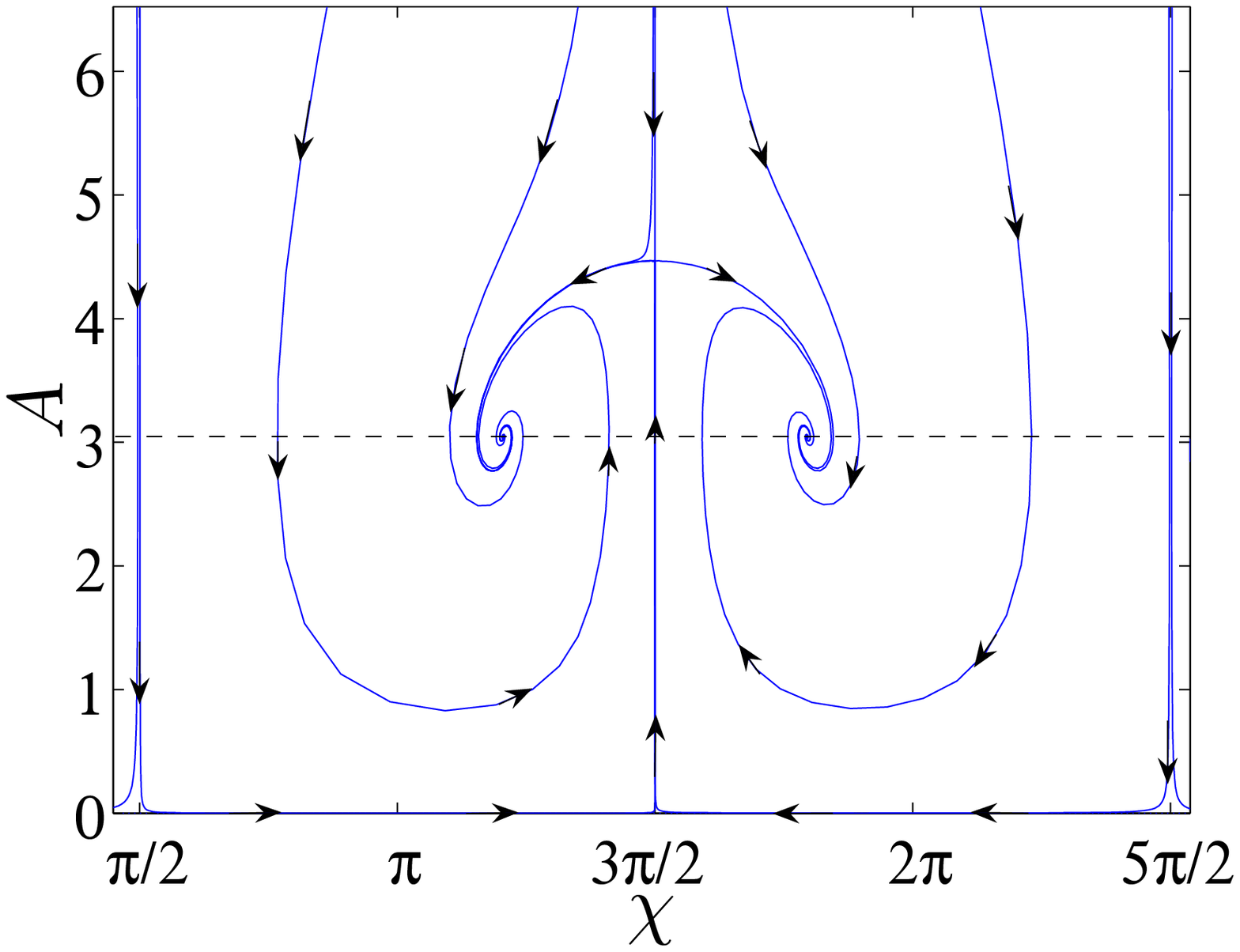}}\hspace*{-.12in}
\subfigure[$\,\,\,\delta=2.5\times10^{-4}$]{\label{finitedelta}\includegraphics[width=0.25\textwidth]{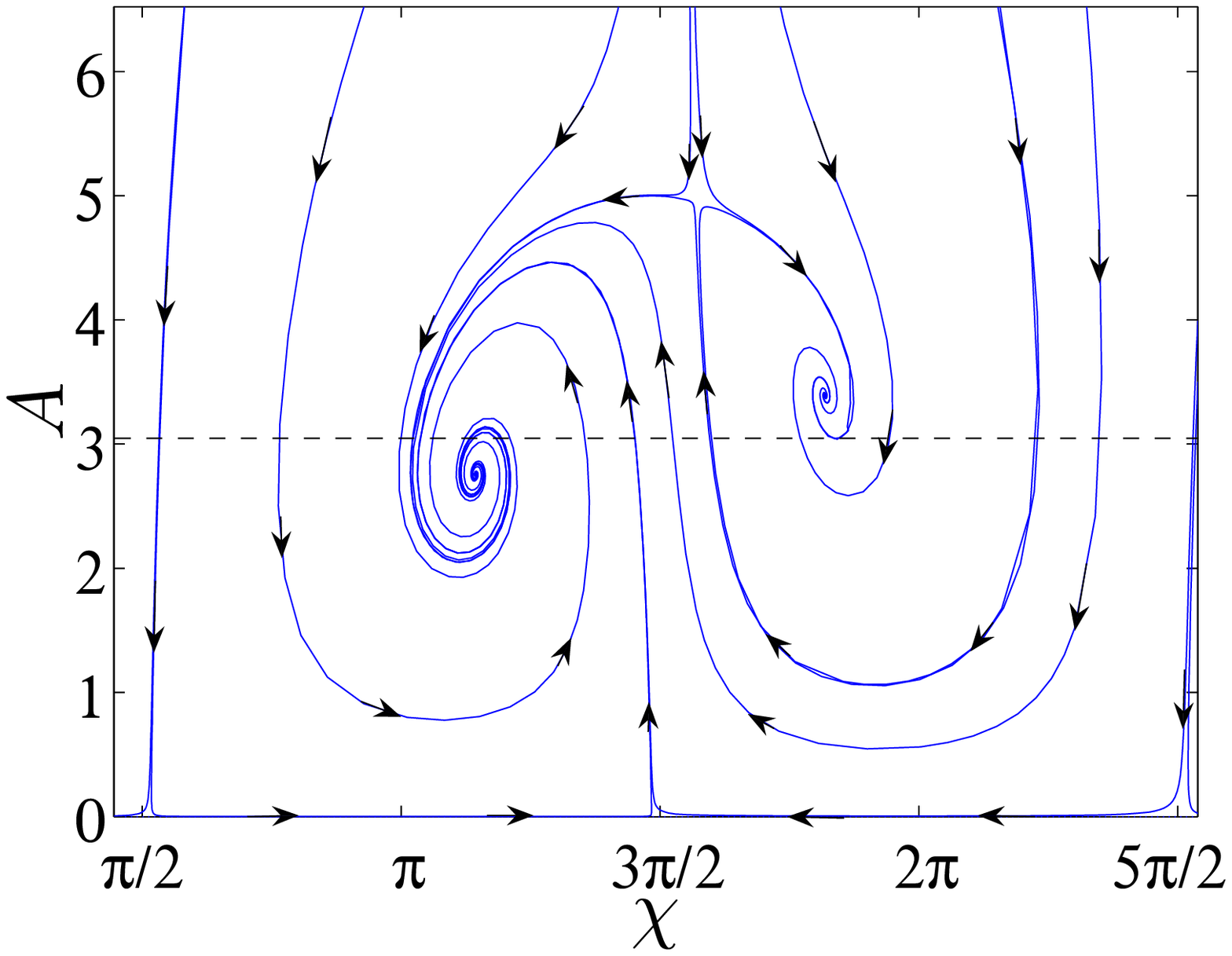}}
\caption{(Color online) Phase space diagrams in the type-$\mathrm{II}$
  regime with $\epsilon=0.01$ and $\tilde\gamma=0.001$ showing (a) two
  asymptotically stable points with vibration amplitude $A_0\sim 3$
  [dashed line] and (b) their shifts from $A_0$ off resonance.}
\label{secondphase}
\end{center}
\end{figure}

Moving off the resonance, $\delta$ becomes non-zero and if
$\epsilon>\epsilon_{\mathrm{II}}$ the degeneracy of the amplitudes
$A_\mathrm{II}^{\pm}$ at the type-$\mathrm{II}$ stationary points is
lifted. If $\delta>0$ the amplitude $A_\mathrm{II}^{+}(\delta)$ is
larger and $A_\mathrm{II}^{-}(\delta)$ smaller than the on-resonance
value $A_{0}$, as shown in Fig.~\ref{secondphase}, while if $\delta<0$
the opposite is true.
As the degeneracy is lifted, the stable type-$\mathrm{II}$ stationary
point that moves to higher amplitudes merges with the
type-$\mathrm{I}$ saddle point and disappears at some critical value
$\pm\delta_{c}$. Consequently, in the interval
$(-\delta_{c},\delta_{c})$ there are two different stable nonlinear
regimes ($\pm$) characterized by different nanotube oscillation
amplitudes and as a consequence by different dc currents through the
system.  A detailed analysis shows that if
$\epsilon-\epsilon_{\mathrm{II}}<<\epsilon_{\mathrm{II}}$ the width
$2\delta_{c}$ of this window of bistability is
$\propto(\epsilon-\epsilon_{\mathrm{II}})^{3/2}$, while the maximum
difference in amplitudes $\vert A_\mathrm{II}^{+}(\pm\delta_{c})-
A_\mathrm{II}^{-}(\pm\delta_{c})\vert$ is
$\propto(\epsilon-\epsilon_{\mathrm{II}})$.

The stationary point that describes the system in a particular
situation depends on the initial conditions. If initially $A\approx
0$, the system always moves to the stationary point with lowest
amplitude as the parametric resonance develops, i.e.
$A_{\mathrm{II}}^{-}$ if $\delta>0$ and $A_{\mathrm{II}}^{+}$ if
$\delta<0$. However, if the system starts from inside the separatrix
defining the higher than on-resonance stationary point (see
Fig.~\ref{secondphase}), it will achieve a stationary amplitude that
is larger than on resonance.  Alternatively, the system can reach this
point if the voltage is slowly changed from resonance, as the system
will follow the trajectory of the stationary point at which it is
defined exactly on resonance. This represents a unique sensitivity in
our system to small changes in the applied bias voltage. Since the dc
current through the system depends on the vibration amplitude,
$j_{dc}\propto A^2$, it follows that we can predict a non-single
valued $I-V$ curve close to resonance. The result is a hysteretic
behavior, the origin of which lies in the multistability of the pumped
nanomechanical vibrations.  This means that the magnitude of the dc
Josephson current in our device is sensitive to the pumping
history. Such memory effects may be employed for different device
applications where the sensitivity of the nanomechanical initial
conditions and the possibility to switch the system between two stable
regimes of vibration can be employed for both sensing and memory
devices. As an example we discuss briefly below how a memory device
could work.
\begin{figure}[h]
\begin{center}
\includegraphics[width=0.45\textwidth]{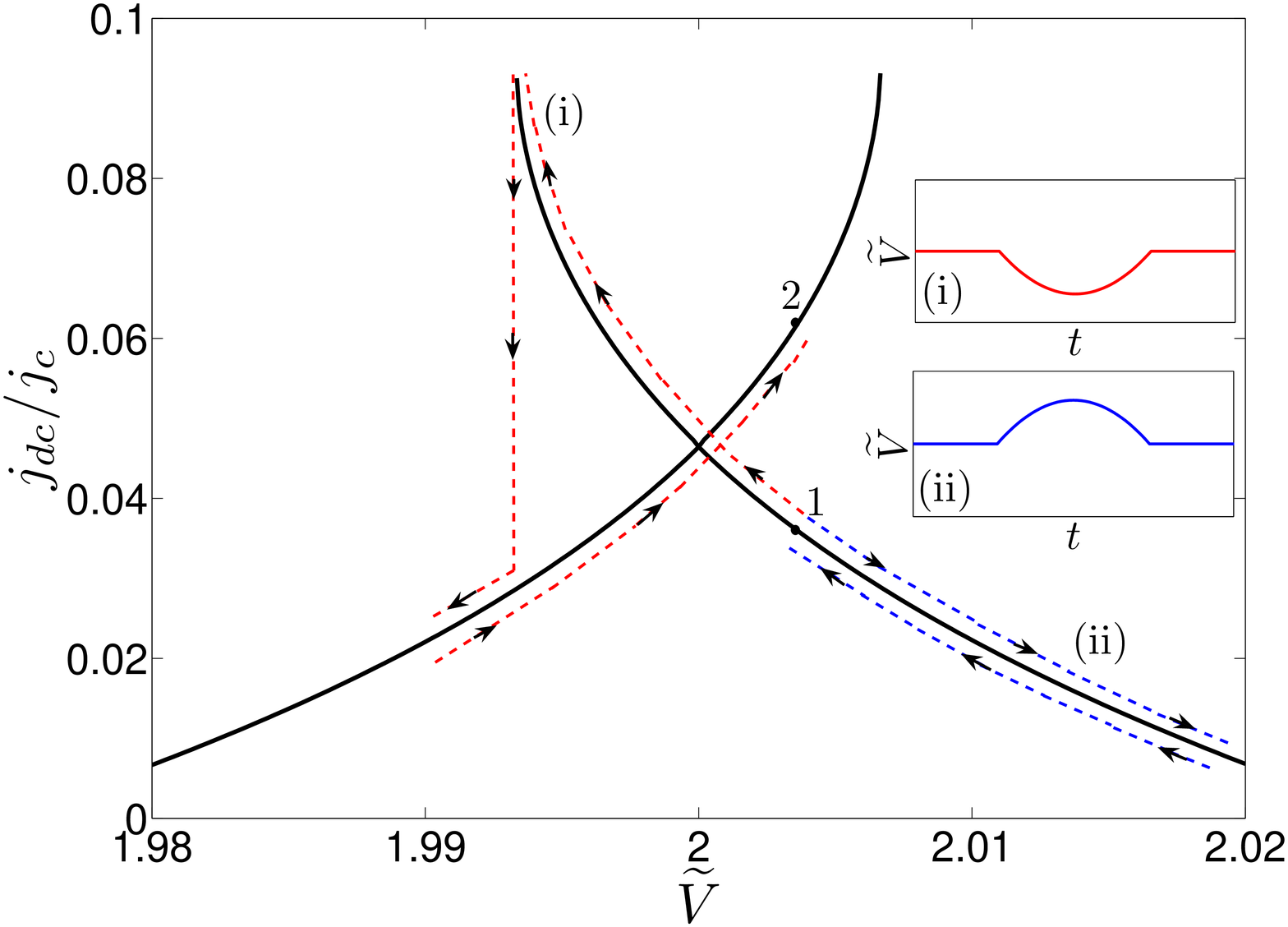}
\caption{(Color online) Diagram illustrating how small voltage pulses
  switch the dc Josephson current. The two pulses in the inset have
  different effects. After pulse (ii) (blue) the current is the same
  as before, while pulse (i) (red) switches the current by a
  measurable amount (from point 1 to 2) thereby storing one bit of
  retrievable information in the device ($\epsilon=0.05$,
  $\tilde\gamma=0.001$).}
\label{currentplot}
\end{center}
\end{figure}

A scheme for the electrical manipulation of our superconducting
nanovibrator is presented in Fig.~\ref{currentplot}, where the
starting position 1 corresponds to a bias voltage which is slightly
off resonance and a nanowire that oscillates with an amplitude
$A_{\mathrm{II}}^{-}(\delta>0)$ smaller than on resonance, see
Fig.~\ref{finitedelta}. Now consider the effect of the voltage pulses
(i) and (ii) shown in the inset.  Pulse (i) moves the system along
trajectory (i) to where the vibration amplitude
$A_{\mathrm{II}}^{-}(\delta<0)$ is larger than on resonance. However,
at $\delta=-\delta_c$ this asymptotically stable point merges with the
third stationary point (saddle) and becomes unstable. The system
therefore jumps to the second asymptotically stable point, where the
vibration amplitude $A_{\mathrm{II}}^{+}(\delta<0)$ is smaller than on
resonance. When the voltage is increased again, the system will move
to position 2, where the vibration amplitude
$A_{\mathrm{II}}^{+}(\delta>0)$ and hence the dc current is larger
than at position 1. One concludes that pulse (i) writes one bit of
information, which is stored as a measurably larger dc current.  Pulse
(ii), on the other hand, moves the initial stability point back and
forth along trajectory (ii) and returns it to the initial position
1. For the parameters considered here, i.e. a resonance frequency of the
order \unit[1]{GHz}, we find that the difference in the current
between points 1 and 2 is a few \unit[]{nA}. Also, the window of
bistability $2\delta_c$ is about \unit[50]{nV} with the second resonance
peak $\widetilde{V}=2$ corresponding to an absolute bias voltage of
$V\sim$ \unit[5]{$\mu$V}. The corresponding mid-point amplitude of
vibration of the nanowire is $\sim$ \unit[25]{nm}.

It is interesting to again compare the phenomena discussed in this
Letter with the Fiske effect \cite{Barone}. Repeating our analysis
we find the low-amplitude behavior to be similar for the two
systems. However, we predict that a dynamical multistability will
appear at a certain value, $\epsilon_{\mathrm{II}}$, of the driving
Lorentz force. This does not occur in the Fiske effect, where the
vibration amplitude exactly on resonance follows the stable solution
corresponding to $\cos\chi=0$ in \eqref{solution2} for all driving
forces $\epsilon$.

To conclude we have shown that for a nanowire suspended between two
voltage-biased superconducting electrodes in a transverse magnetic
field, pronounced resonance phenomena can be found at discrete values
of the driving voltage. Our analysis shows that the behavior of this
system is governed by an effective equation of motion whose solution
gives the amplitude of the nanowire oscillations and the dc Josephson
current as a function of system parameters.  Most importantly, it was
shown that for realistic experimental parameters the system can be
driven into a multistable regime by varying the magnetic field strength.
The possibility to pump energy into the mechanical vibrations of a
suspended nanowire and the ensuing dynamical multistability of the
vibration amplitude and dc current makes this superconducting
nanoelectromechanical device a unique system, where the sensitivity to
initial conditions and switching between two stable regimes can be
probed experimentally.

Financial support from the Swedish VR and SSF is gratefully
acknowledged.
\bibliography{/chalmers/users/sonneg/Desktop/Artiklar/referencesdriven}
\end{document}